\documentclass[twocolumn,english,prl,amssymb,aps,superscriptaddress,showpacs,twocolumn,amsmath,showkeys,floatfix]{revtex4-1}
\usepackage[T1]{fontenc}
\usepackage[latin9]{inputenc}
\usepackage{geometry}
\usepackage[active]{srcltx}
\usepackage{textcomp}
\usepackage{amsmath}
\usepackage{graphicx}
\usepackage{color}
\usepackage{esint}
\usepackage{svg}
\usepackage{siunitx}
\usepackage{physics}
\usepackage[version=4]{mhchem}
\makeatletter
\usepackage{babel}
\newcommand{\im}{\mathrm{i}}
\newcommand{\e}{\mathrm{e}}

\makeatother

\begin{document}
	
	\title {Modulation transfer protocol for Rydberg RF receivers}
	\author{Duc-Anh Trinh}
	\affiliation{Universit\'e Paris-Saclay, CNRS, ENS Paris-Saclay, CentraleSup\'elec, LuMIn, 91190 Gif-sur-Yvette, France}
	\author{Adwaith K.V.}
	\affiliation{Universit\'e Paris-Saclay, CNRS, ENS Paris-Saclay, CentraleSup\'elec, LuMIn, 91190 Gif-sur-Yvette, France}
	\author{Mickael Branco}
	\affiliation{Universit\'e Paris-Saclay, CNRS, ENS Paris-Saclay, CentraleSup\'elec, LuMIn, 91190 Gif-sur-Yvette, France}
	\author{Ali\'enor Rouxel}
	\affiliation{Universit\'e Paris-Saclay, CNRS, ENS Paris-Saclay, CentraleSup\'elec, LuMIn, 91190 Gif-sur-Yvette, France}
	\author{Sacha Welinski}
	\affiliation {Thales Research and Technology, 91120 Palaiseau, France}
	\author{Perrine Berger}
	\affiliation {Thales Research and Technology, 91120 Palaiseau, France}
	\author{Fabienne Goldfarb}
	\affiliation{Universit\'e Paris-Saclay, CNRS, ENS Paris-Saclay, CentraleSup\'elec, LuMIn, 91190 Gif-sur-Yvette, France}
	\author{Fabien Bretenaker}
	\affiliation{Universit\'e Paris-Saclay, CNRS, ENS Paris-Saclay, CentraleSup\'elec, LuMIn, 91190 Gif-sur-Yvette, France}

	\begin{abstract} 
We propose and demonstrate a modulation transfer protocol to increase the detection sensitivity of a Rydberg RF receiver to fields out of resonance from the transition between Rydberg levels. This protocol is based on a phase modulation of the control field used to create the Electromagnetically Induced Transparency (EIT) signal. The nonlinear wave-mixing of the multi-component coupling laser and the probe laser transfers the modulation to the probe laser, which is used for RF-field detection. The measurements compare well with semi-classical simulations of atom-light interaction and show an improvement in the RF bandwidth of the sensor and an improved sensitivity of the response to weak fields.
	\end{abstract}
	
	
	\maketitle
	
Research on RF receivers based on optical detection methods using thermal Rydberg atoms has strongly developed in the last decade. From a three-level EIT scheme in Rubidium 87 (\ce{^{87}Rb}) vapor, this RF detection method based on the induced Autler-Townes splitting of Rydberg levels has been expanded to Rubidium 85 (\ce{^{85}Rb}), Cesium (Cs) vapors \cite{Sedlacek2012,Fan2015,Kumar2017,6910267} and four-level EIT schemes \cite{Brown2023,Chen2022,Thaicharoen2019,Bohaichuk2023}. It is capable of detecting the signal from the sub-Gigahertz range \cite{Brown2023} to Terahertz frequencies \cite{Chen2022} using the numerous allowed atomic transitions between Rydberg states in alkali atoms \cite{Chopinaud2021}. The large electric dipole moments of these transitions make Rydberg RF receivers very sensitive to resonant microwave fields, with sensitivities in the range of \si{\uV / \m} in the \si{\GHz} frequency range. However, the signal quickly decreases when the detuning of the RF signal from the atomic transition increases. An RF reference field serving as a local oscillator can help retrieve more sensitivity \cite{Simons2019,Jing2020,Yao2022}, but this method is more cumbersome and requires an auxiliary antenna, which prevents the system from being purely dielectric and can, therefore, induce RF field distortion.

We propose an all-optical method based on a modulation transfer by four-wave mixing in the atomic medium \cite{Ducloy1982}, which improves the sensitivity of the atom-based RF receiver to detuned fields. Although such a modulation transfer method is popular for laser frequency stabilization using two-level systems \cite{MTS2008,Preuschoff2018,Sun2016}, there are very few experimental demonstrations of this technique applied to three- or four-level systems \cite{Lim2022}. Here, we consider such a modulation transfer method in the case of a three-level ladder EIT phenomenon in \ce{^{85}Rb} and investigate its capability of increasing the detection bandwidth of the corresponding Rydberg RF receiver. 

In the present letter, we thus implement both the conventional and the modulation transfer protocols and compare their responses to RF fields with different detunings. We focus on the low-field domain where the Autler-Townes (AT) splitting is hardly visible, and we interpret the experimental results with numerical simulation.

\begin{figure}[tbp]
\includegraphics[width = \linewidth]{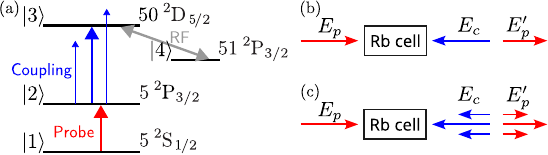}
\caption{\label{fig1}(a) Rubidium levels used for the EIT-based RF receiver. (b) Conventional protocol, with single-frequency coupling and output probe fields. (c) Modulation transfer protocol, with sidebands for the coupling and output probe fields.}
\end{figure}

We use \ce{^{85}Rb}, which has the highest abundance in a natural mixture of 85 and 87 isotopes. A probe laser around 780 \si{\nm} couples its ground state $\ket{1}=$ [\ce{5^2S_{1/2}}($F = 3;\, m_F$)] to the intermediate state $\ket{2}=$ [\ce{5^2P_{3/2}}($F=4;\,m_F$)] (Fig.\,\ref{fig1}. (a)). A 480 \si{\nm} laser couples this intermediate state manifold to the Rydberg state $\ket{3}=$ [\ce{50^2D_{5/2}}($F=3,4,5;\,m_F$)]. We couple this Rydberg state to the state $\ket{4}=$ [\ce{51^2P_{3/2}}($F=2,3,4;\,m_F$)] with the RF field around 17.0422 \si{\GHz} that we wish to detetct. 

In the conventional protocol, the probe and coupling laser beams are single-frequency fields and can be modeled by 
$
E_{p,c} = \mathcal{E}_{p,c} \e^{-\im \omega_{p,c} t} + 
\qcc*
$ (see Fig. \ref{fig1}. (b)).
The transmitted probe beam after the vapor cell then remains single-frequency and can be written
$
E_{p}' = \mathcal{E}_{p}' \e^{-\im \omega_{p,c} t} + 
\qcc*
$ 

In the modulation transfer protocol, the coupling laser is phase-modulated and exhibits sidebands with opposite signs (see Fig. \ref{fig1}. (c)):
\begin{eqnarray}
\label{eq1}
E_c &=& -\mathcal{E}_{c_{1}} \e^{-\im (\omega_c -\omega_{mod}) t} + 
\mathcal{E}_{c_{0}} \e^{-\im \omega_c t} \nonumber \\
&& \> + \mathcal{E}_{c_{1}} \e^{-\im (\omega_c +\omega_{mod}) t} + \qcc*,
\end{eqnarray}
where $\omega_{mod}$  is the modulation frequency. Close to EIT resonance, the modulation is transferred from the coupling beam to the probe beam via the atomic medium nonlinearities so that one now expects the transmitted probe laser output to also exhibit sidebands at $\omega_{mod}$ from the carrier:
\begin{eqnarray}
\label{eq2}
E_p' &=& \mathcal{E}_{p_{-1}}' \e^{-\im (\omega_p -\omega_{mod}) t} + 
\mathcal{E}_{p_{0}}' \e^{-\im \omega_p t} \nonumber \\
&& \> + \mathcal{E}_{p_{+1}}' \e^{-\im (\omega_p +\omega_{mod}) t} + \qcc*
\end{eqnarray}
The response of the modulation transfer protocol is based on the beat notes of the carrier of $E_p'$ and its radiated sidebands, which results from four-wave mixing in Rubidium vapor. This nonlinear response differs from the linear response achieved by lock-in detection employing frequency modulation of the probe laser \cite{Kumar2017a} or chopping of the coupling laser \cite{Sedlacek2012}.  

\begin{figure}[htbp]
\includegraphics[width = \linewidth]{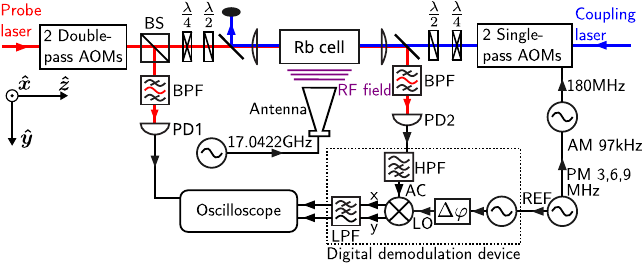}
\caption{\label{fig2}Schematized experimental setup. BS - beam splitter, $\frac{\lambda}{4}$ - quarter-wave plate, $\frac{\lambda}{2}$ - half-wave plate, BPF - bandpass filter, PD - photodetector, \textcolor{black}{HPF - highpass filter, AM - amplitude modulation}, PM - phase modulation, AC - alternating current, \textcolor{black}{REF - reference}, $\Delta \varphi$ - phase shifter, LO - local oscillator, \textcolor{black}{x,y - two orthogonal quadratures}, LPF - lowpass filter. Photodetectors 1 (PD1) and 2 (PD2) monitor the probe laser before and after the vapor cell, respectively. In the conventional protocol, \textcolor{black}{the coupling laser is amplitude-modulated by a square signal}. In the modulation transfer protocol, \textcolor{black}{the coupling laser is phase-modulated by a sinusoidal signal. A copy of the modulation signal is sent to the demodulation device as the reference.}
}
\end{figure}

Figure \ref{fig2} schematizes the experimental setup used to test this idea. The probe and coupling lasers counter-propagate through a quartz cell containing a natural mixture of Rubidium (with 73\% of \ce{^{85}Rb} and 27\% of \ce{^{87}Rb}) at room temperature. Inside the cell, the probe (coupling) beam has a waist diameter of 0.3 (0.4) \si{\mm} with an 11.5 (14.5) \si{\cm} Rayleigh length, which is longer than the 7.5 \si{\cm} vapor cell length. The probe input power is 0.4 \si{\uW}, and the coupling one is \textcolor{black}{46} \si{\mW}. A horn antenna (MVG QR18000) is placed 57.5 \si{\cm} away from the vapor cell and emits an RF field perpendicularly to the laser path. RF absorbing panels isolate the antenna and the vapor cell from other elements. The probe, coupling, and RF fields are linearly polarized along $\vu*{x}$, so that all the magnetic sub-levels between the two Rydberg states are coupled \cite{Sedlacek2013}. 

We use a Toptica tunable diode laser to provide the 780 \si{\nm} probe laser and a Toptica amplified and frequency-doubled tunable diode laser (TA-SHG) system to generate the 480 \si{\nm} coupling laser. The probe laser is stabilized by Modulation Transfer Spectroscopy \cite{MTS2008} on the atomic frequency $\omega_{21}$ that corresponds to the transition between the levels $\ket{1}$ and $\ket{2}$. Using an auxiliary cell, the coupling laser is then locked using the EIT resonance in the ladder system \cite{Abel2009}, which enforces the condition $\omega_p + \omega_c = \omega_{31}$, where $\omega_{31}$ is the frequency difference between the atomic levels $\ket{1}$ and $\ket{3}$. Considering the RF frequency value, which gives a symmetrical AT splitting, we obtain $\omega_{34}/2\pi\simeq17.0422$ \si{\GHz} for the RF transition between the levels $\ket{3}$ and $\ket{4}$. The RF signal is provided by an R\&S SMF100A signal generator, and the RF electric field amplitude $E_{RF}$ at the vapor cell position is estimated after taking into account cable and insertion losses, the gain of the antenna and its distance to the vapor cell. The cell perturbation factor \cite{Robinson2021a}, which decreases the average RF field inside the cell, is estimated to be about \textcolor{black}{0.52} from a finite-difference time-domain (FDTD) electromagnetic simulation around 17 \si{\GHz} with our vapor cell and experimental characterization using AT splitting.

\begin{figure}[tbp]
\includegraphics[width = \linewidth]{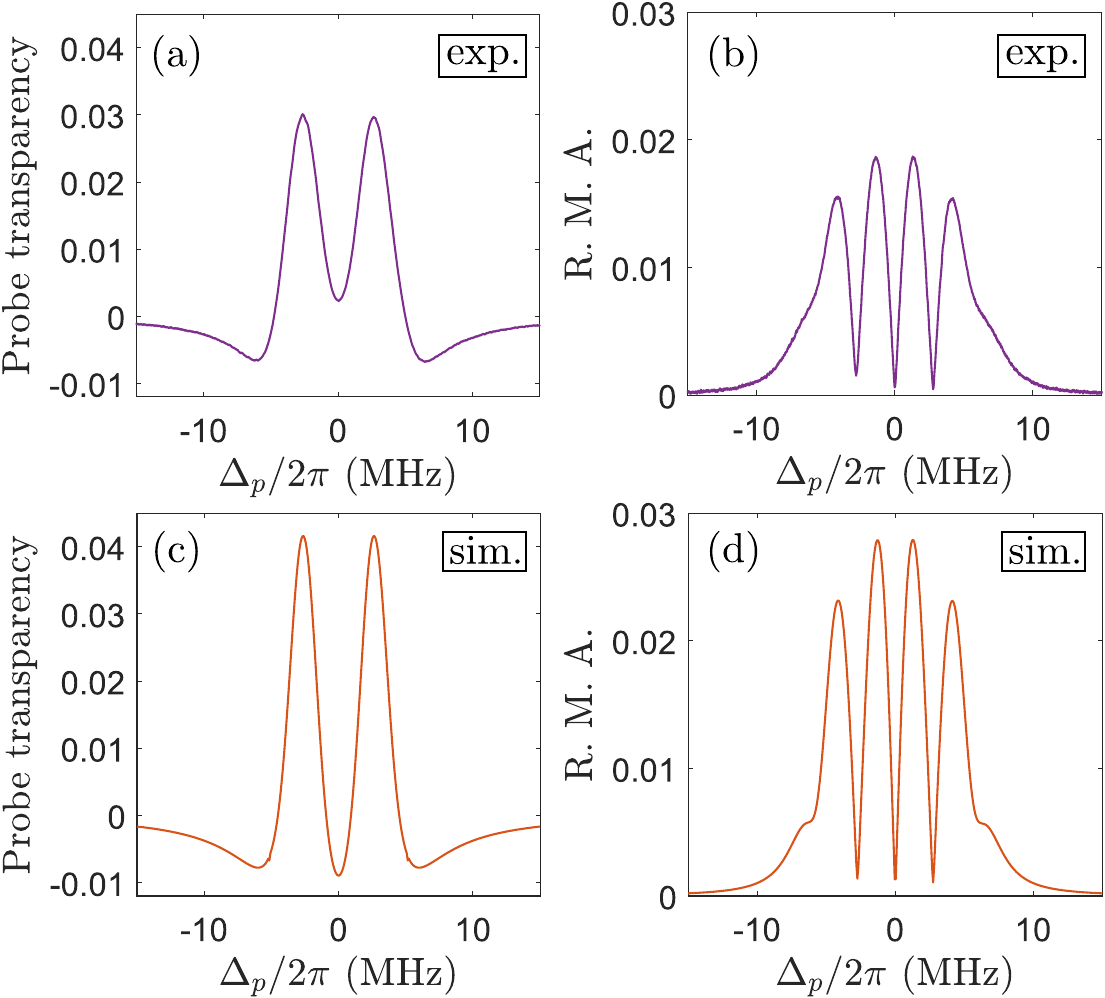}
\caption{\label{fig3}Experimental spectra of RF receiver response for (a) the conventional protocol and (b) the modulation transfer protocol at modulation frequency $\omega_{mod}/ 2 \pi = 3$ \si{\MHz}. The RF electric field amplitude is at \textcolor{black}{$E_{RF} = 0.5014$ \si{\V / \m}}, and the RF frequency is at $\omega_{RF}/ 2\pi = 17.0422$ \si{\GHz}. The coupling laser is at resonance with the transition between $\ket{2}$ and $\ket{3}$. Corresponding simulation results for (c) conventional and (d) modulation transfer protocols.}
\end{figure} 

Let us now examine the signal obtained with the conventional and modulation transfer protocols as a function of probe laser detuning $\Delta_p = \omega_p - \omega_{21}$. We scan the probe laser frequency around $\omega_{21}$ by $\pm$ 100 \si{\MHz} thanks to two double-pass acousto-optic modulators (AOM) \cite{Donley2005}. \textcolor{black}{While the first AOM is driven by a fixed frequency from a signal generator, the second AOM is controlled by a VCO, with a 10 \si{\Hz} scanning rate. We modulate the coupling beam differently in each protocol. The coupling beam travels through two single-pass AOMs. The first one shifts the light frequency by -180 \si{MHz}, whereas the second one shifts it back by 180 \si{\MHz} and encodes the modulations. In the conventional protocol, we perform lock-in detection to enhance signal readout. Then, the coupling beam is amplitude-modulated with a 97-\si{\kHz} square signal of 50\% duty-cycle.} 
A fast APD (Thorlabs APD410A/M) of 10 \si{MHz} instantaneous bandwidth and $1.2 \times 10^7$ \si{V/W} sensitivity at 780 \si{nm} detects the probe laser after the vapor cell. \textcolor{black}{The detected signal is demodulated by a lock-in amplifier device (Zurich UHF 600 \si{\MHz}). With a proper demodulation phase, one quadrature drops to zero, and the other one gives the difference of the transmitted probe beam between the on- and off-states of the coupling beam. Another photodetector monitors the probe laser before the vapor cell so that we can record the corresponding input power}. We average the received signal over 128 scanning rounds to reduce electronic noise. Knowing the photodetector sensitivity and the electronic attenuation, we can derive the \textcolor{black}{probe laser transparency ($\abs{\mathcal{E}_p'(\text{coupling on})}^2 / \abs{\mathcal{E}_p}^2 - \abs{\mathcal{E}_p'(\text{coupling off})}^2 / \abs{\mathcal{E}_p}^2$) from the probe input and the maximized quadrature. One example of such a transparency signal in the presence of a resonant RF field is plotted in Fig. \ref{fig3}(a), together with the result of the corresponding simulation in Fig. \ref{fig3}(c).}

These simulations are obtained by the density matrix formalism using an effective four-level atomic system \cite{Holloway2017}. This effective model is reasonable because every hyperfine magnetic-sublevel ($\ket{F \, m_F}$) in our polarization configuration participates in a four-level excitation established by $\vu*{x}$-polarized probe, coupling, and RF fields. Rabi frequencies are derived considering the effective electric dipole elements for linear transitions between Zeeman sublevels ($m_F \rightarrow m_F' = m_F$). The electric dipole elements and decay rates for the first transition ($\ket{1} \rightarrow \ket{2}$) are taken from well-established data \cite{Steck2023}. For the second ($\ket{2} \rightarrow \ket{3}$) and third ($\ket{3} \rightarrow \ket{4}$) transitions, we obtain these parameters from an open-source Python package \cite{Sibalic2017,Robertson2021}. To take into account the atomic medium optical thickness \cite{Rotunno2023} (linear transmission equal to \textcolor{black}{0.37}), we divide the cell into a hundred layers, for which the thin medium approximation can be considered valid. We include the Doppler effect of the thermal vapor by integrating over velocity classes and the transit rate of atoms through the laser beam by corresponding decay and feeding rates \cite{Finkelstein2023}. The simulation results, shown in Figs. 3(c) and (d), are in good agreement with the measurements for both protocols, except concerning the amplitude of the signal in presence of the RF field. This discrepancy is due to the simplified 4-level system considered for the simulations, which does not take into account the broadening due to the different AT splittings for different transitions between the many different sublevels of the system.
In order to simulate the modulation transfer protocol, a Floquet expansion and the continued fraction method \cite{Papademetriou1992,Wong2004} allow handling the oscillating terms in the optical Bloch equations. 

Now, to implement the modulation transfer protocol experimentally, we need to modulate the phase of the coupling beam. \textcolor{black}{By applying a phase modulation at a frequency $\omega_{mod}/2\pi \ll 180$ \si{\MHz} to the driving signal of the second AOM, this phase-modulation is imprinted on the coupling light \cite{Li2005}.} The coupling laser thus exhibits two opposite sidebands at $\pm \omega_{mod}$ from the carrier (see Eq. \ref{eq1}). Thanks to the 16 \si{\MHz} phase-modulation bandwidth of the AOM and the 10 \si{\MHz} instantaneous bandwidth of the photodetector, we could test several phase modulation frequencies $\omega_{mod} / 2 \pi = 3,\, 6,\, 9$ \si{\MHz}. The phase modulation amplitude of the RF signal injected into the AOM is $\pi / 3$. A scanning Fabry-Perot interferometer allowed to measure a 0.6 ratio $\abs{\mathcal{E}_{c_{\pm 1}} / \mathcal{E}_c }$ between the amplitudes of the sidebands and the carrier. 
\textcolor{black}{Once the transmitted probe beam and its intensity modulation are detected, we demodulate the signal at $\omega_{mod}$ and filter the two quadratures, which reconstruct the amplitude of the beat note. From this and the probe input, we derive the relative modulation amplitude (R.M.A.), which is defined as $2 
\abs{\mathcal{E}_{p_{-1}}' \mathcal{E}_{p_0}'^* + 
\mathcal{E}_{p_0}' \mathcal{E}_{p_{+1}}'^*
} / \abs{\mathcal{E}_{p}}^2
$.} 
Figure \ref{fig3}(b) shows a typical RMA signal evolution versus probe detuning in the presence of the RF field. The signal of Fig. \ref{fig3}(b) now contains many more features than the usual DC signal of Fig. \ref{fig3}(a), which we will exploit in the following. The shape ot the RMA spectrum is the result of interferences between the beat notes created by the beating of the transferred sidebands $\mathcal{E}_{p_{-1}}'$ and $\mathcal{E}_{p_{+1}}'$ with the carrier  $\mathcal{E}_{p_{0}}'$. The out-of-phase beat notes interfere destructively, turning the maxima and minima of Fig. \ref{fig3}(a) into narrow dips. The dual-peak spectrum of Fig. \ref{fig3}(a) is thus replaced by three very sharp dips in Fig. \ref{fig3}(b). The sharpness of these dips is at the origin of the improved sensitivity to changes in the RF field amplitude, which will be discussed later. Notice also that the shape of the spectrum of Fig. \ref{fig3}(b) is confirmed by the simulations of Fig.  \ref{fig3}(d).

\begin{figure}[tbp]
\includegraphics[width = \linewidth]{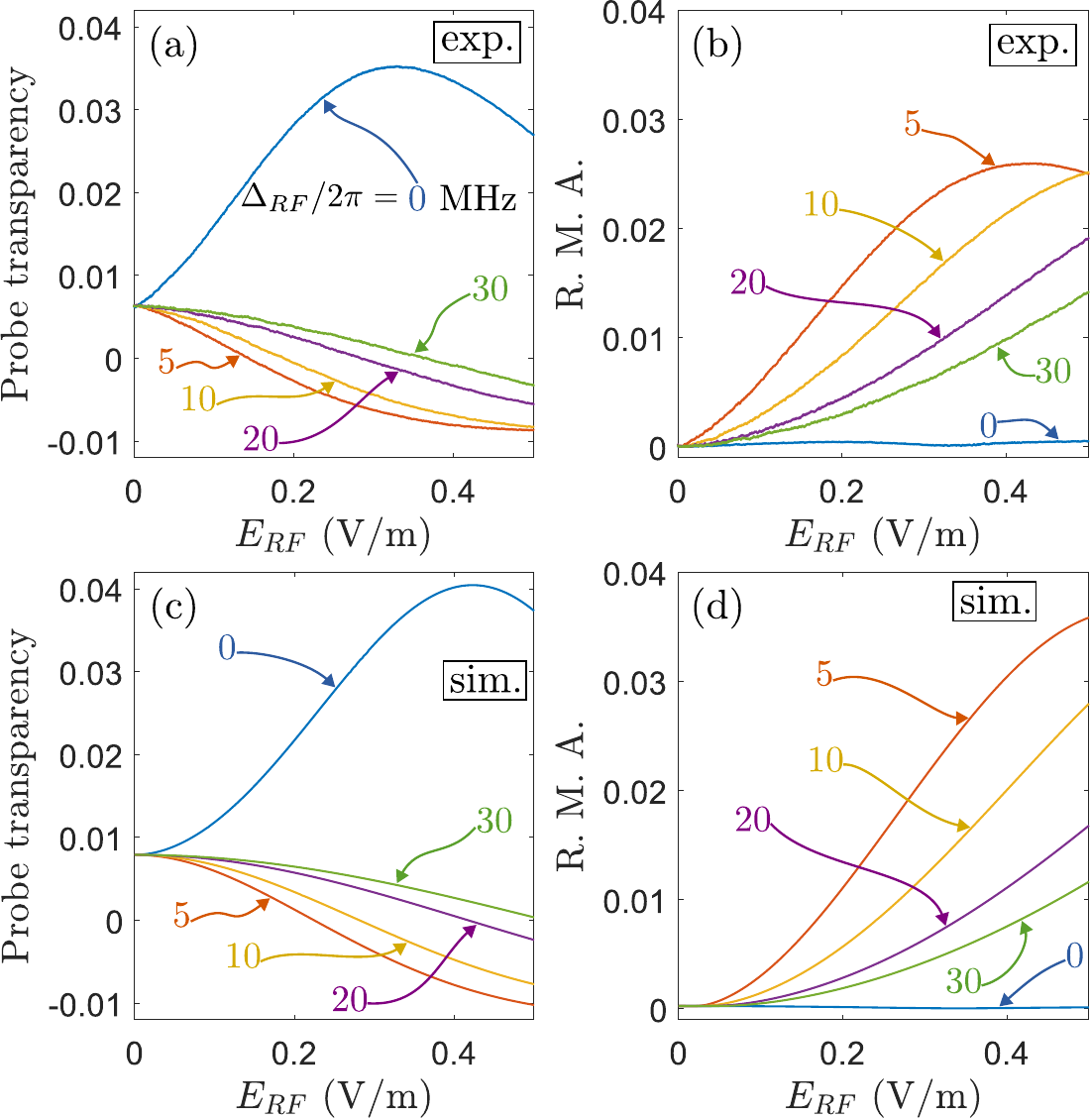}
\caption{\label{fig4} \textcolor{black}{Recorded signal as a function of electric field amplitude ($E_{RF}$) for different RF detunings ($\Delta_{RF}$). (a) Experiment: conventional protocol at $\Delta_p / 2 \pi = 2$ \si{\MHz}. (b) Experiment: modulation transfer protocol at $\Delta_p / 2 \pi = 0$ \si{\MHz} and $\omega_{mod} / 2 \pi$ = 3 \si{\MHz}. (c, d) Corresponding simulation results.}}
\end{figure}

Let us now investigate the system response to RF fields at different RF detunings $\Delta_{RF} = \omega_{RF} - \omega_{34}$. In the following, for both protocols, the probe laser detuning $\Delta_p=\omega_p-\omega_{21}$ is fixed, and the RF electric field amplitude $E_{RF}$ is scanned at a rate of 10 \si{\Hz}. A fast oscilloscope characterizes the amplitude-modulated RF field to ensure the modulation parameters provide a linear scan. Different sets of data were recorded for $\Delta_ p / 2 \pi = 0, \pm 2, \pm 4$ \si{\MHz}, and for $\Delta_{RF} / 2 \pi = 0, 5, 10, 20, 30\,\mathrm{MHz}$. We choose to reproduce here those corresponding to the probe detuning, which gives the best sensitivity for detuned RF fields ($\Delta_{RF} > 0$) for each protocol. Figures \ref{fig4}(a) and (b) thus show the results obtained for $\Delta_ p / 2 \pi = 2$ \si{\MHz} for the conventional protocol and $\Delta_ p / 2 \pi = 0$ \si{\MHz} for the modulation transfer one. At $\Delta_p / 2\pi = 0$, two out-of-phase components in the probe beat note for the resonant RF field result in a flat response ($\Delta_{RF} / 2\pi = 0$ \si{\MHz} curve, Fig. \ref{fig4}(b)). The RF detuning causes an unbalance between these two components and gives rise to the RF response. The corresponding simulations (Figs. \ref{fig4}(c) and (d)) predict the behavior of the system. The remaining differences between the simulation and experiment originate from the simplification introduced in the simulated model by using an effective four-level system. Many sub-AT-splittings from RF magnetic transitions ($\ket{F\,m_F} \rightarrow \ket{F'\,m_{F'}}$) lead to stronger peak broadening than that in the simulation. This explains why the maxima in the experimental curves (Fig. \ref{fig4}(a), (b)) happen at smaller values of $E_{RF}$ than in the simulated ones (Fig. \ref{fig4}(c), (d)).

When the RF field is at resonance, it is clear that the conventional protocol gives the best sensitivity to variations of the RF field amplitude: the slope of the blue curve of Fig. \ref{fig4}(a) is very large. However, as soon as the RF field is detuned by a few \si{\MHz}, the response of this signal to variations of $E_{RF}$ becomes very flat (see the curves for $\Delta_{RF}\neq0$ in Fig. \ref{fig4}(a)), indicating a poor sensitivity of the standard protocol to detuned RF field variations. On the contrary, the signal obtained from the modulation transfer protocol (see the curves for $\Delta_{RF}\neq0$ in Fig. \ref{fig4}(b)) exhibits a strong slope for RF detunings even as large as 30 \si{\MHz}. \textcolor{black}{To assess their performance quantitatively, we fit the signal responses by polynomials and calculate their derivatives to obtain the signal slope amplitudes. }  


\begin{figure}[h!]
\includegraphics[width = 0.8\linewidth]{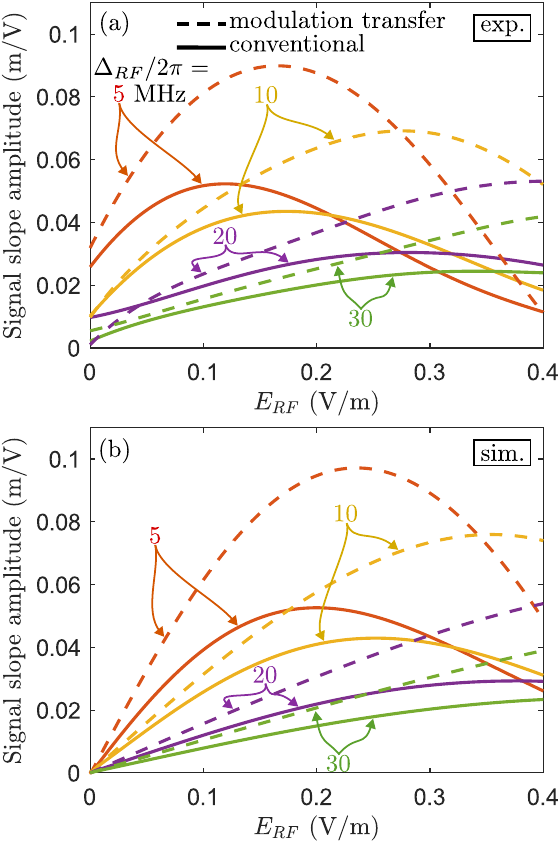}
\caption{\label{fig5} (a) Experimental signal slope versus RF field amplitude for different RF field detunings. Full lines: conventional DC signal. Dashed line: Modulation transfer protocol. (b) Simulated signal slope versus RF field amplitude, 
}
\end{figure}

Figure \ref{fig5} compares such slopes for different RF detunings for both protocols. Assuming that the detection noise is shot noise limited, the sensitivity $\mathcal{S}$ of the system is derived from the slopes of Fig. \ref{fig5} via 
\begin{eqnarray} \color{black}
\label{eq3}
\mathcal{S} = \frac{\sqrt{2e}}{\dv{T}{E_{RF}} \sqrt{P \eta}},
\end{eqnarray}

where $T$ is the probe transmission, $e$  the electron charge, $P$  the transmitted probe power, $\eta$  the sensitivity of the photodetector, and $\dv{T}{E_{RF}}$  the slope extracted from the data of Fig. \ref{fig5}. \textcolor{black}{Considering $P$ as constant}, the sensitivity improvement due to the modulation protocol depends only on the slope improvement. \textcolor{black}{The experimental slopes, reproduced in Fig.\,\ref{fig5}(a), show that the modulation transfer protocol exhibits a better response up to $E_{RF} = 0.4$ \si{\V / \m}. One exception is for the RF detuning at 20 \si{\MHz}, where the modulation transfer slope is lower than the conventional one at very low fields $E_{RF} < 0.05$ \si{\V / \m}. We believe that this is due to the electromagnetically noisy environment, which modifies the signal response at very low RF fields. The simulated slopes reproduced in Fig. \ref{fig5}(b) support the claim that the modulation transfer method is always better at very low fields. Some examples of sensitivity improvements from the experimental results are as follows, around $E_{RF} = 0.05$ \si{\V / \m}, $\mathcal{S}$ is increased by about 10\% for $\Delta_{RF} / 2\pi = 10,\, 20,$ \si{\MHz}, 20\% for $\Delta_{RF} / 2\pi = 30$ \si{\MHz}, and 40\% for $\Delta_{RF} / 2\pi = 5$ \si{\MHz}. The improvement reaches up to 100\% at around $E_{RF} = 0.3$ \si{\V / \m} for $\Delta_{RF} / 2\pi = 5,\, 10$ \si{\MHz}. For this value of $E_{RF}$, it reaches 40\% for $\Delta_{RF} / 2\pi = 30$ \si{\MHz}, and 50\% for $\Delta_{RF} / 2\pi = 20$ \si{\MHz}. 
}

If the detection is shot noise limited, Eq.\,(\ref{eq3}) leads to sensitivities of the order of 0.1 \si{\uV  \cm^{-1} \Hz^{-1/2}} for the standard protocol when the RF field is at resonance. Using the modulation protocol, we find the same sensitivity for an RF field detuned by 10 \si{\MHz}. Such sensitivities compare favorably with the existing literature \cite{Kumar2017, Kumar2017a}.

In this article, we have thus proposed and demonstrated an RF detection method using Rydberg-EIT in the presence of phase modulation of the coupling beam. We have shown theoretically and experimentally that this modulation is transferred to the probe beam in the presence of EIT, and that, once demodulated, the probe modulation can improve the sensitivity of an EIT-based Rydberg RF receiver to weak and detuned fields. With our parameters, the transfer of modulation to the probe provides a new response to the RF field, leading to a higher sensitivity for small RF fields observed for detunings as large $\Delta_{RF}/2\pi = 30$ \si{\MHz}. 

One future development of the present work consists of building detection electronics able to record the signals from the two protocols simultaneously to take advantage of high resonant RF sensitivity from the conventional one and the improved non-resonant RF sensitivity from the modulation transfer one. Modulation parameters can be adapted to different use cases, corresponding to other alkali atoms or other Rydberg levels. Other approaches to extend this protocol further include tailoring a constant sensitivity using a quadrature with a well-chosen modulation configuration, testing different modulation depths, and implementing a sophisticated modulation and demodulation system to explore different nonlinear responses.

We thank Joseph Delpy and Antoine Browaeys for fruitful discussions and S\'ebastien Rousselot for technical support.

This work was funded by the French Defense Innovation Agency, Quantum Saclay, and the European Defence Fund (EDF) under grant agreement EDF-2021-DIS-RDIS-ADEQUADE (n$^{\circ}$101103417). 

Funded by the European Union. Views and opinions expressed are however those of the author(s) only and do not necessarily reflect those of the European Union or the European Commission. Neither the European Union nor the granting authority can be held responsible for them.

\section*{DATA AVAILABILITY}
The data that support the findings of this study are available from the corresponding author upon reasonable request.
%

\end{document}